\documentclass[twocolumn]{aastex61}
\received{August 28, 2018}
\revised{October 10, 2018}
\accepted{October 11, 2018}
\submitjournal{Astronomical Journal}

\shortauthors{Moran et al.}

\begin{document}

\title{Limits on Clouds and Hazes for the TRAPPIST-1 Planets}

\correspondingauthor{Sarah E. Moran}
\email{moran.sarahe@gmail.com}

\author{Sarah E. Moran}
\affil{Department of Earth and Planetary Sciences,
Johns Hopkins University,
Baltimore, MD 21218, USA}

\author{Sarah M. H\"{o}rst}
\affiliation{Department of Earth and Planetary Sciences,
Johns Hopkins University,
Baltimore, MD 21218, USA}

\author{Natasha E. Batalha}
\affiliation{Space Telescope Science Institute,
Baltimore, MD 21218, USA}
\affiliation{Department of Astronomy and Astrophysics, University of California, Santa Cruz, CA 95064, USA}

\author{Nikole K. Lewis}
\affiliation{Department of Earth and Planetary Sciences,
Johns Hopkins University,
Baltimore, MD 21218, USA}
\affiliation{Space Telescope Science Institute,
Baltimore, MD 21218, USA}
\affiliation{Department of Astronomy and Carl Sagan Institute, Cornell University, 122 Sciences Drive, Ithaca, NY 14853, USA}

\author{Hannah R. Wakeford}
\affiliation{Space Telescope Science Institute,
Baltimore, MD 21218, USA}



\begin{abstract}

The TRAPPIST-1 planetary system is an excellent candidate for study of the evolution and habitability of M-dwarf hosted planets. Transmission spectroscopy observations performed on the system with the \emph{Hubble Space Telescope (HST)} suggest that the innermost five planets do not possess clear hydrogen atmospheres. Here we reassess these conclusions with recently updated mass constraints. Additionally, we expand the analysis to include limits on metallicity, cloud top pressure, and the strength of haze scattering.  We connect recent laboratory results of particle size and production rate for exoplanet hazes to a one-dimensional atmospheric model for TRAPPIST-1 transmission spectra. In this way, we obtain a physically-based estimate of haze scattering cross sections. We find haze scattering cross sections on the order of 10$^{-26}$ to 10$^{-19}$ cm$^2$ are needed in modeled hydrogen-rich atmospheres for TRAPPIST-1 d, e, and f to match the \emph{HST} data. For TRAPPIST-1 g, we cannot rule out a clear hydrogen-rich atmosphere. We modeled the effects an opaque cloud deck and substantial heavy element content have on the transmission spectra using the updated mass estimates. We determine that hydrogen-rich atmospheres with high altitude clouds, at pressures of 12 mbar and lower, are consistent with the \emph{HST} observations for TRAPPIST-1 d and e. For TRAPPIST-1 f and g, we cannot rule out clear hydrogen-rich cases to high confidence. We demonstrate that metallicities of at least 60$\times$ solar with tropospheric (0.1 bar) clouds are in agreement with observations. Additionally, we provide estimates of the precision necessary for future observations to disentangle degeneracies in cloud top pressure and metallicity. For TRAPPIST-1 e and f, for example, 20 ppm precision is needed to distinguish between a clear atmosphere and an atmosphere with a thick cloud layer at 0.1 bar across a wide range (1$\times$ to 1000$\times$ solar) of metallicity. Our results suggest secondary, volatile-rich atmospheres for the outer TRAPPIST-1 planets d, e, and f.

\end{abstract}




\section{Introduction} \label{sec:intro}
Aerosols, including clouds and hazes, in the atmospheres of exoplanets are currently the subject of intense scrutiny. These aerosols can hinder our ability to study the presence and composition of exo-atmospheres, and their presence is often invoked to explain a lack of large spectral features in transmission spectroscopy studies \citep[e.g.,][]{knutson2014featureless,knutson2014hd978,kreidberg2014clouds,dragomir2015rayleigh,sing2016continuum}. Yet, much remains uncertain as to the likelihood of aerosol formation and physical properties in unexplored radiation regimes, such as those of M-dwarf systems like the TRAPPIST-1 planets. Here, we connect \emph{Hubble Space Telescope (HST)} observations to a laboratory-based study of exoplanetary haze properties to investigate the likelihood of aerosols in hydrogen-rich atmospheres for the TRAPPIST-1 system. 

The  TRAPPIST-1  system \citep{gillon2017seven} is  the  first known multi-planet system of Earth-sized worlds. Additionally, three to four of its currently known planets are in the classically defined habitable zone, and all seven currently known are amenable to observational transmission spectroscopy studies. Here and throughout this work, we refer to the ``classical'' habitable zone, defined as the circumstellar region where liquid water can persist on a planet's surface given a substantial planetary atmosphere \citep{kasting1993habitable}.  As such, TRAPPIST-1 is a powerful natural system for insights into the formation and evolution of planetary atmospheres in a system outside our own. Furthermore, the \emph{Transiting Exoplanet Survey Satellite (TESS)}, launched in April 2018, is expected to find hundreds of terrestrial planets around M-dwarf stars in the nearby galactic neighborhood \citep{sullivan2015tess}. These new worlds will have their own unique radiation regimes and will require further study to understand their aerosol content and the effects of any aerosols on future observations. 

Previous studies of the TRAPPIST-1 planets have investigated their orbital evolutionary histories \citep[]{luger2017seven,quarles2017plausible,tamayo2017convergent,unterborn2018inward}. These planets have long-term stable orbits, which provides adequate time for substantial evolution of the planetary atmospheres \citep{dong2018atmospheric}.  Despite the high UV-flux of the host star \citep[]{omalleyjames2017uv,wheatley2017strong} as well as frequent flaring events \citep{vida2018flares}, the planets of the TRAPPIST-1 system still may have large amounts of water \citep[]{,Bolmont2017water,bourrier2017temporal}. This has motivated multiple investigations of the planets' habitability through the presence of surface liquid water and biomarkers \citep[]{barstow2016habitable,alberti2017comparative,wolf2017assessing,turbet2018modeling}. Studies regarding the interior structure \citep[]{kislyakova2017magma,suissa2018trappist} and bulk densities \citep{grimm2018nature} of the TRAPPIST-1 planets also suggest terrestrial rather than gaseous worlds. 

Observations  from \emph{HST}  have  determined  that  TRAPPIST-1 d, e, and f have muted transmission spectra with features in the $<$500 ppm range, rather than the large features ($\sim$1000 ppm) expected for extended, clear hydrogen-rich atmospheres \citep{dewit2018atmospheric}. ``Hydrogen-rich'', in this case and throughout our analysis, refers to H$_2$-He envelopes greater than 0.01\% of the total planet masses given their radii.  \citet{dewit2018atmospheric} ruled out clear hydrogen-rich atmospheres for TRAPPIST-1 d, e, and f to high confidence. Later, the TRAPPIST-1 planetary mass measurements were updated \citep{grimm2018nature}. The scale height of the atmosphere, given by $H = \frac{kT}{\mu g}$ where \emph{k} is the Boltzmann constant, \emph{T} is temperature, $\mu$ is the mean molecular weight of the atmosphere, and \emph{g} is the gravity, is dependent on planetary mass. The predicted atmospheric scale heights for these planets have thus changed in light of these new mass estimates. Therefore, the initial findings of \citet{dewit2018atmospheric}, in which clear hydrogen-rich atmospheres were ruled out, must be revisited. We do so as part of our analysis for this study. Any further mass refinements will have an effect on these results, which we explore in more detail in \S4.

Clouds and/or hazes in planetary atmospheres can obscure and mute the larger spectral features indicative of clear hydrogen-rich atmospheres. There is ample evidence for the presence of clouds and hazes in the atmospheres of planets across all masses, from hot Jupiters \citep[e.g.,][]{sing2016continuum} to exo-Neptunes \citep[e.g.,][]{knutson2014featureless,lothringer2018hst} to super-Earths \citep[e.g.,][]{kreidberg2014clouds}. Hazes in these atmospheres are of particular astrobiological interest. Hazes can substantially impact the planetary surface temperature as well as provide a source of UV absorption to protect the planetary surface \citep[e.g.,][]{arney2017pale}. TRAPPIST-1 A has high flux in the  UV and X-ray \citep{wheatley2017strong}, as is typical of late M-dwarfs \citep{france2013ultraviolet}. Protection from the high flux of the host star would likely be paramount for any possible life to persist. Furthermore, hazes themselves are thought to be important for the formation of prebiotic molecules \citep[e.g,][]{horst2012formation,horst2018neoproterozoic,rimmer2018rna}.

The formation mechanisms and physical likelihoods of clouds and hazes in the atmospheres of exoplanets, and their subsequent effect on observations, remain largely unknown. Nearly all previous studies of haze and clouds in exoplanet atmospheres have depended upon the particle sizes and compositions of Solar System photochemical hazes \citep[e.g.,][]{howe2012theoretical,millerriccikempton2012chemistry,morley2013quantatively,morley2015thermal,rackham2017optical,lincowski2018trappist} rather than those in exoplanet atmospheres, as no direct measurements of these aerosols are currently possible. From these Solar System-like hazes, previous studies then assume these properties \emph{a priori} and provide predictions for the types of transmission and emission spectra expected. 

Here, we take the opposite approach, by making no direct assumptions about the cloud or haze species contributing to these transmission spectra. Instead, we calculate from the recent \emph{HST} \citep{dewit2018atmospheric} observations upper limits on possible cloud and hazes, remaining agnostic about their origin. We then compare these values to recent experimental work investigating exoplanet haze properties for the first time in the laboratory. These exoplanet experiments studied haze formation under a range of planetary temperatures and atmospheric compositions, including under hydrogen-rich, water-rich, and carbon dioxide-rich cases \citep{he2018laboratory,he2018uvhaze,horst2018haze}. These experiments are thus applicable to expected scenarios for the TRAPPIST-1 planets near and within the classical habitable zone. As such, our investigation represents a new approach to characterizing the TRAPPIST-1 planets.

In \S 2, we describe our methodology for modeling spectra informed by the laboratory results; in \S 3, we present our results; in \S 4, we discuss and contextualize the results we have obtained; and in \S 5 we summarize our findings with concluding remarks.

\section{Methods} \label{sec:method}
We aim to determine lower limits on the cloud top pressures and upper limits on the strength of haze scattering in the outer TRAPPIST-1 planets d, e, f, and g. To do so, we use \emph{CaltecH Inverse ModEling and Retrieval Algorithms} (or \texttt{CHIMERA}) \citep{line2013chimera} to generate model transmission spectra to compare to the \emph{HST} observations of these atmospheres. Table \ref{table:planetparams} shows the TRAPPIST-1 planetary parameters explored here, with the measurements we used in our models. We begin by first noting an important distinction between our definitions of a cloud versus a haze, as these terms are often used interchangeably when, at least in our analysis, we use them to mean very specific physical phenomena. Hazes refer to the solid, suspended particles that are the result of photochemical reactions in the atmosphere. Clouds, on the other hand, are either solid or liquid particles that are the result of condensation processes due to temperature and pressure conditions of the atmosphere. Together, these two sets of suspended particles are encompassed by the generic term ``aerosols''.

We investigate three different effects clouds and hazes can have on an atmosphere. First, for each planet, we investigate the effect of adding haze into our spectral models. We fix the composition of a cloud-free model while varying the magnitude of Rayleigh-scattering haze. We increase the strength of the haze until the model attains a statistically significant agreement to the \emph{HST} data, using a cutoff threshold of reduced-$\chi^2$ of 1.16 (or 1$\sigma$, based on the 10 \emph{HST} data points) and then again until we reach a reduced $\chi^2$ of 2.8 (3$\sigma$). If we cannot reach these cutoffs, we report the highest confidence value we are able to obtain, with all of our results, in Table \ref{table:results}.

Second, we examine the effect of atmospheric composition in a cloudy atmosphere. We fix the strength of haze scattering to zero and keep the cloud top pressure at 0.1 bar, the expected pressure of the tropopause \citep{robinson2014tropopause}, while varying the amount of water in the atmosphere. We increase the water mixing ratio until our model reaches the 1$\sigma$ and 3$\sigma$ uncertainty bounds of the \emph{HST} data. Third, we fix the atmospheric composition and vary the cloud-top pressure, effectively moving the cloud layer in altitude until we again reach the 1$\sigma$ and 3$\sigma$ thresholds. The simplicity of this method is motivated by the low precision of the \citet{dewit2018atmospheric} data. We describe each of these methods in more detail in the following subsections.
    
\subsection{Modeling Haze Opacity with Laboratory Measurements}
Different methods exist to account for the effect of hazes on transmission spectra in exoplanet atmospheres. These techniques range in complexity from including the output of a full set of haze species' opacity coupled to photochemical models \citep[e.g.,][]{millerriccikempton2012chemistry,morley2013quantatively} to a more simplistic treatment wherein the haze scattering is parameterized by a power law \citep[e.g.,][]{robinson2012model,line2013chimera,line2014systematic,robinson2014tropopause,sing2016continuum}. This simpler method is often employed where data are not sufficiently precise to merit more complex treatment.

In light of the large uncertainties associated with the \emph{HST} TRAPPIST-1 observations \citep{dewit2018atmospheric}, we use the parameterization for the scattering cross section, $\sigma$, derived in \citet{lecavelier2008rayleigh}:
\begin{equation} \label{eq:scattering}
    \sigma = \sigma_0(\lambda/\lambda_0)^\alpha
\end{equation}
where $\sigma_0$ is the reference scattering cross section, $\lambda_0$ is a reference wavelength, $\lambda$ is the wavelength of radiation, and $\alpha$ is the power law slope of scattering. The Rayleigh approximation, in which $\alpha$ $=$ -4, applies when the diameter of particle d$_p$~$<<$~$\lambda$. The exoplanet haze analogues from the laboratory range in particle size \emph{d$_p$} from 25 nm to 180 nm \citep{he2018laboratory,he2018uvhaze}. For the wavelengths covered by the \emph{HST}/WFC3 observations, from $\lambda$ = 1.1 $\mu$m to 1.7 \micron, this allows us to treat these exoplanet haze analogues as Rayleigh scatterers in our model. The scaling factor of the haze cross section $\sigma_0$ can be approximated by 
\begin{equation} \label{eq:cross section}
    \sigma_0 = \frac{\tau_0}{n_0H}
\end{equation}
where $\tau_0$ is the optical depth at a reference altitude, \emph{n$_0$} is the number density of the scatterer at the reference altitude, and \emph{H} is the scale height of the atmosphere.  We compute upper limits on $\sigma_0$ based on the \emph{HST} observations and then compare these results to the laboratory results of \citet{horst2018haze} to determine whether our computed upper limits are physically plausible. 

There is no direct physically motivated formulation to turn the exoplanet haze production efficiencies from the laboratory \citep{horst2018haze} to a theoretical scattering cross section. However, laboratory results from \citet{horst2018haze} show maximum production rates for 300 K to 600 K exoplanetary atmospheres similar to the production rate of Titan experiments with the same experimental set-up. Therefore, we are able to use Titan as a benchmark because its haze formation is well-studied in the laboratory \citep[e.g.,][]{he2017carbon} as well as directly observed through remote sensing. 

To approximate the connection from laboratory production rates to Eqn. \ref{eq:scattering}, we calculate $\sigma_0$ as defined in Eqn. \ref{eq:cross section} from the combined results of \citet{tomasko2008model} and \citet{robinson2014titan}. \citet{tomasko2008model} reports the measured haze particle number density as a function of altitude on Titan, and \citet{robinson2014titan} used the haze parametrization of \citet{lecavelier2008rayleigh} to fit the slope of Titan solar occultation observations, in effect treating Titan as an exoplanet in transit. \citet{tomasko2008model} and \citet{robinson2014titan} suggest a scattering cross section for Titan's haze of $\sigma_0$ $\sim$ $10^{-7}$~cm$^2$. For reference, Earth's scattering cross section is on the order of $\sim10^{-27}$~cm$^{2}$. The haze production rates for the exoplanet experiments were not higher than for similar Titan experiments \citep{horst2018haze}. Therefore, we assume that this scaled haze scattering cross section 10$^{-7}$ cm$^2$ is a reasonable physical upper limit for haze scattering in the TRAPPIST-1 atmospheres.

We reiterate the important caveat to our methodology described here: the production efficiency measured in the laboratory does not directly correspond to the cross sectional strength of the haze observed in a planetary atmosphere. The laboratory cannot capture processes such as atmospheric escape and rainout, which would both work to decrease the column density and decrease the cross sectional scattering strength. Thus our assumption of the production efficiency being representative of the column density of the haze particles is an approximation. For our purposes of attempting to estimate the haze content of the TRAPPIST-1 planets, our approximation is justified by the large uncertainties already inherent to the \emph{HST} observations from \citet{dewit2018atmospheric}, as well as guided by our overall goal of obtaining an upper limit rather than an exact constraint of the atmospheric scattering due to haze.  

\subsection{Modeling Cloud Opacity}

There are many techniques to account for the effects of clouds on transmission spectra. These techniques exist along a continuum of complexity. These range from full 3-D dynamically-radiatively-convectively driven cloud microphysics models \citep[e.g.,][]{lee2015modelling}, to models which globally average the balance between turbulent mixing and sedimentation of condensates \citep[e.g.,][]{ackermanmarley2001}, to simply modeling a grey opacity source \citep[e.g.,][]{line2013chimera,batalha2017information}. Here, we choose to use the method of \citet{batalha2018strategies}, where a grey absorbing cloud is set at a specific pressure, below which the transmittance is zero. This allows us to remain agnostic about the properties of the cloud being formed in each case, and provides a lower limit on where an optically thick, global cloud layer would have to exist to match the \emph{HST} observations. This method has been used in observations of hot Jupiters \citep[e.g.,][]{sing2016continuum}, warm Neptunes \citep[e.g.,][]{kreidberg2014clouds,wakeford2017hatp26b}, and studies of the TRAPPIST-1 system \citep[e.g.,][]{morley2017observing, batalha2018strategies}. 

To provide additional context for our cloudy cases, we also run models wherein we place this opaque cloud at the nominal tropopause of 0.1 bar. This follows the analysis of \citet{robinson2014tropopause}, which observes that all Solar System planetary bodies have a tropopause at 0.1 bar, where thick clouds are observed to form.  In our models with tropospheric clouds, we examine the effect of increasing the metallicity of the atmospheres by varying the water mixing ratio. Multiple studies have suggested that the outer TRAPPIST-1 planets in the classical habitable zone, due to their likely origin further out in the protostellar disk before inward migration \citep[]{quarles2017plausible,unterborn2018inward}, are still able to harbor multiple Earth ocean’s worth of water despite having lost huge amounts of water over their evolutionary histories \citep[]{Bolmont2017water,bourrier2017temporal,dong2018atmospheric}. Recent revisions of mass estimates for the planets \citep{grimm2018nature} allows a large water reservoir to remain a reasonable assumption \citep{unterborn2018updated}. This motivates our approach to using water as our proxy for heavy atmospheric enrichment.  We also focus on water because of the significant water feature at 1.4 $\micron$ centered in the \emph{HST}/WFC3 bandpass. Additionally the equilibrium temperatures of the outer TRAPPIST-1 planets d, e, f, and g, ranging from about 300 K to 200 K, include water's triple point. Water could then contribute to atmospheric dynamics on the TRAPPIST-1 planets as it does on Earth through cloud condensation and rainout processes \citep[e.g.,][]{turbet2018modeling}. We start from the solar water mixing ratio of 7.8 x 10$^{-4}$ (with a solar C/O ratio of 0.5) \citep{lodderssolar2003}, and increase this value until reaching our statistical thresholds of 1$\sigma$ and 3$\sigma$.

Finally, for clouds we explore the degeneracy between cloud top pressure and metallicity by running a full grid of forward models in fixed parameter combinations. We examine a range of metallicities from 1$\times$ to 1000$\times$ solar (in fifteen logarithmic steps), as well as a range of cloud top pressures (in ten steps) from a clear atmosphere (i.e., where molecular opacity becomes optically thick well before the cloud deck) to a cloud deck at 1 $\mu$bar. Our grid thus includes a total of 150 distinct models each for TRAPPIST-1 d, e, f, and g.

\subsection{Modeling the Transmission Spectra}
We use a version of the atmospheric modeling code \texttt{CHIMERA} \citep{line2013chimera,line2013nearir,line2014systematic}, a radiative transfer code that uses the correlated-$k$ distribution technique. \texttt{CHIMERA} has been used to model hot Jupiters \citep{kreidberg2014hotjupiter}, sub-Neptunes \citep{kreidberg2014clouds} and recently the TRAPPIST-1 system \citep{batalha2018strategies}. Given molecular opacities, planetary mass, radius, temperature, atmospheric mixing ratios, cloud top pressure level, and haze cross section ($\sigma$), we produce transmission spectra at R = 100, consistent with the \textit{HST}/WFC3 measurements. 

For our temperature-pressure profiles, we use the parameterized 1-D, 5 parameter profile of \citet{guillot2010TPprofile}. We include molecular opacities from to H$_2$/He CIA, CH$_4$, H$_2$O, CO$_2$, and N$_2$ \citep{freedman2008opacities,freedman2014opacities}, informed by the dominant gases in the atmospheres of Solar System worlds as well as previous TRAPPIST-1 atmospheric studies \citep[]{morley2017observing,batalha2018strategies}. For our atmospheric mixing ratios, we take two separate approaches: one composition for hazy atmospheres and a separate set of compositions for cloudy atmospheres. Both of our approaches involve setting mixing ratios rather than exploring a fully self-consistent calculation of gases in chemical equilibrium. The TRAPPIST-1 atmospheres are likely not in chemical equilibrium \citep{dong2018atmospheric, bourrier2017temporal}. Additionally, in our models with photochemical haze formation, we inherently assume that this is not the case. We assign mixing ratios in order to determine upper limits on aerosol content rather than providing constraints for any physical atmospheric parameters. 

For our hazy atmospheres, we use H$_2$/He background gas atmospheres with 1\% mixing ratios of CH$_4$, H$_2$O, CO$_2$, and N$_2$, giving a mean molecular weight $\mu$ of 3.02. This composition is motivated by two factors. First, nitrogen, methane, water, and other carbon-bearing species have all been shown to play important roles in haze production in laboratory settings \citep[e.g.,][]{imanaka2010nitrogen,trainer2012nitrogen,horst2014co,trainer2004haze}. Second, these constituent gases were used in the relevant exoplanet laboratory haze experiments \citep[]{he2018laboratory,he2018uvhaze,horst2018haze}.

For our cloudy atmospheres, we consider atmospheres of H$_2$/He and H$_2$O since we are comparing our results to the \emph{HST}/WFC3 G141 bandpass, where water has a prominent molecular feature. Therefore we use the water mixing ratio as a proxy for varying the scale height due to increasing metallicity (where metallicity refers to the overall heavy-element abundance). We begin from a H$_2$/He atmosphere with a solar H$_2$O mixing ratio with a mean molecular weight $\mu$ of 2.32 and then scale upward to higher metallicities by increasing the water mixing ratio.

\begin{table}[h!]
\centering
\begin{tabular}{||c c c c||} 
 \hline
 Planet & Mass ($\earth$) & Radius ($\earth$) & T\textsubscript{eq} (K) \\ [0.5ex] 
 \hline\hline
 d & 0.297 & 0.784 & 288.0 \\ 
 e & 0.772 & 0.910 & 251.3 \\
 f & 0.934 & 1.046 & 219.0 \\
 g & 1.148 & 1.154 & 198.6 \\ [1ex]
 \hline
\end{tabular}
\caption{TRAPPIST-1 planet parameters via \citet{grimm2018nature} used in our model atmospheres.}
\label{table:planetparams}
\end{table}

\section{Results} \label{sec:floats}

\begin{figure*}[!h]
\centering
\includegraphics[angle=0,width=0.95\linewidth]{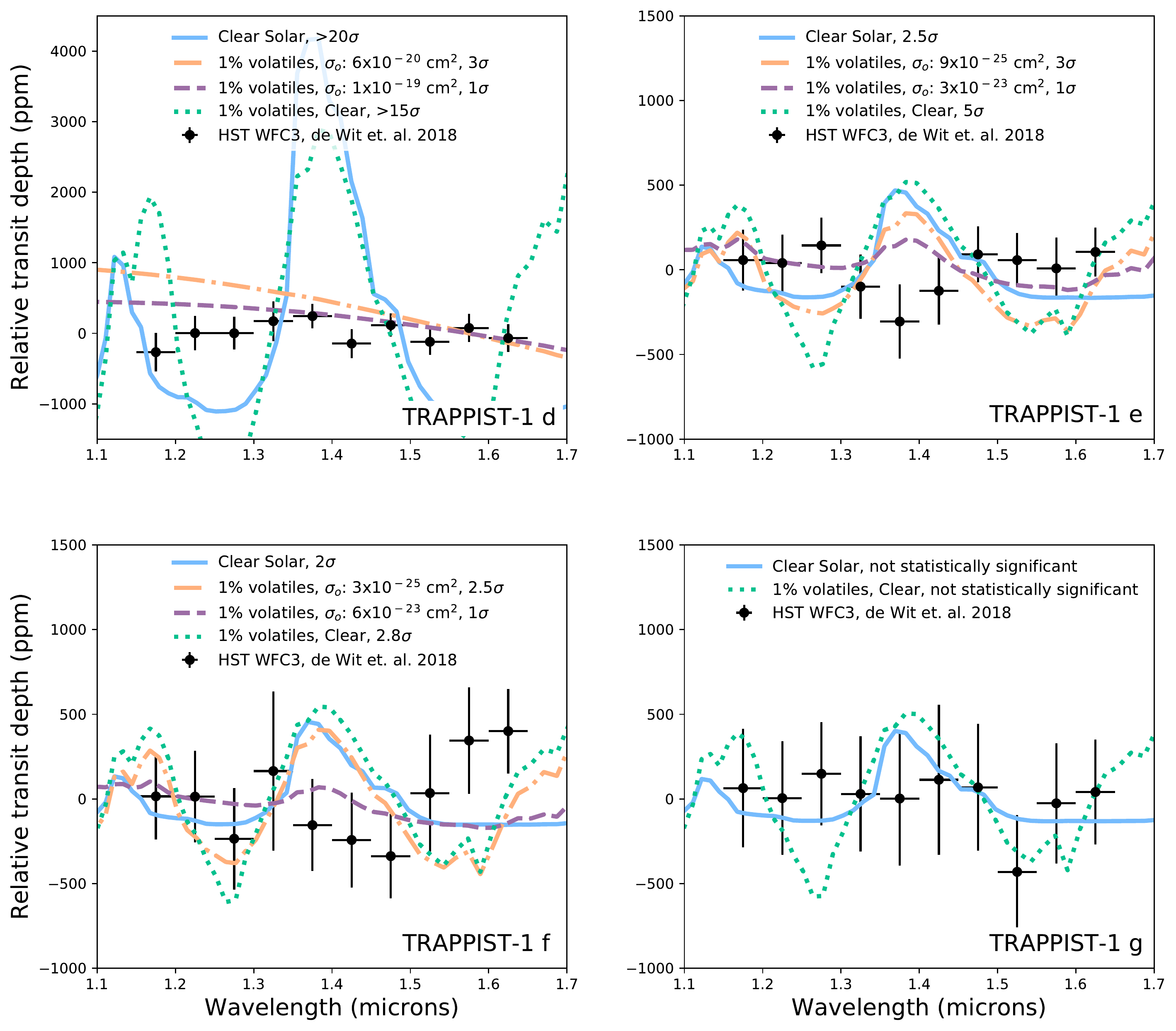}
\caption{Black circles and error bars indicate the previous \emph{HST}/WFC3 observations \citep{dewit2018atmospheric} for TRAPPIST-1 planets d, e, f, and g. The blue solid lines indicate our baseline aerosol-free solar composition case. The rest are models of a hydrogen-rich atmosphere containing Rayleigh scattering haze, with mixing ratios for water, carbon dioxide, nitrogen, and methane at 1\%. Green dotted lines indicate a zero magnitude haze scattering cross section; purple dashed lines indicate that the haze cross section was increased to give 1$\sigma$ agreement with the \emph{HST} data; orange dash-dot lines display haze cross sections increased to give 3$\sigma$ agreement with the \emph{HST} data. Only planet d results in the clear exclusion of a haze-free atmosphere.}
\label{fig:haze}
\end{figure*}

Photochemical hazes are not expected to persist in temperate hydrogen-rich atmospheres from theory \citep[]{hu2014photochemistry,millerriccikempton2012chemistry,morley2015thermal}. Laboratory measurements \citep{he2018laboratory,horst2018haze} suggest inefficient photochemical haze production for hydrogen-rich background gas mixtures. Our results in Figure \ref{fig:haze} show the model outputs for hydrogen-rich atmospheres containing 1\% volatiles with haze amplitudes to the 1$\sigma$ and 3$\sigma$ confidence levels for \emph{HST} data of TRAPPIST-1 d, e, and f. Clear hydrogen-rich atmospheres with 1\% volatiles as well as solar composition atmospheres are also shown for planets d, e, f, and g. Additionally, we show in Figure \ref{fig:clouds} our model outputs to 1$\sigma$ and 3$\sigma$ levels for a global cloud deck at the nominal tropopause with high metallicity atmospheres as well as for a global cloud deck with a solar composition atmosphere. Finally, we compare the results of increasing metallicity to increasing cloud top pressure in Figure \ref{fig:statistics}. We show that aerosols, either as photochemical haze or as an opaque equilibrium cloud layer, are likely unable to mute spectral features to within the observational uncertainties from \citet{dewit2018atmospheric} for all but planet g, if we consider the laboratory-supported haze production rates. A summary of our results given the model conditions and the statistical significance of the models is found in Table \ref{table:results}.
\begin{table}[h!]
\centering
\begin{tabular} {|p{1cm}||p{2cm}p{3cm}|}
\hline
Planet & \multicolumn{2}{|c|}{Haze Scattering Cross Section (cm$^2$)}\\
\hline
& 1$\sigma$ & 3$\sigma$ \\
\hline\hline
d & 1$\times$10$^{-19}$ & 6$\times$10$^{-20}$   \\
e & 3$\times$10$^{-23}$ & 9$\times$10$^{-25}$   \\
f & 6$\times$10$^{-23}$ & 3$\times$10$^{-25}$ (2.5$\sigma$) \\
g & $<1\sigma$ & $<1\sigma$ \\
\hline\hline\hline
\end{tabular}

\begin{tabular}{|p{1cm}||p{2cm}p{3cm}|}   
\hline
Planet & \multicolumn{2}{|c|}{Cloud Top Pressure (bar)}\\
\hline
& 1$\sigma$ & 3$\sigma$\\
\hline\hline
d & 8$\times$10\textsuperscript{-7} & 2$\times$10\textsuperscript{-6} \\
e & 2$\times$10\textsuperscript{-2} & 1$\times$10\textsuperscript{-1} (2$\sigma$) \\
f & 1.26$\times$10\textsuperscript{-2} & clear (2$\sigma$) \\
g & $<1\sigma$ & $<1\sigma$ \\
\hline\hline\hline
\end{tabular}

\begin{tabular}{|p{1cm}||p{2cm}p{3cm}|}  
\hline
Planet & \multicolumn{2}{|c|}{Metallicity ($\times$solar)}\\
\hline
& 1$\sigma$ & 3$\sigma$ \\
\hline\hline
d & 500 & 300 \\
e & 500 & 100 \\
f & 630 & 60 \\
g & $<1\sigma$ & $<1\sigma$ \\
\hline
\end{tabular}
\caption{Summary of upper and lower limits found from model outputs for our test cases with statistical certainties to \emph{HST} data. Our haze scattering cross sections represent the scattering strength needed to reach 1$\sigma$ and 3$\sigma$ agreement to the \emph{HST} data in hydrogen/helium atmospheres with 1\% H$_2$O, CO$_2$, CH$_4$, and N$_2$ mixing ratios. Cloud top pressures given are the lower boundary of pressure levels required in each atmosphere with a solar composition to agree with the \emph{HST} data to 1$\sigma$ and 3$\sigma$. The metallicity given is the lower limit of the water mixing ratio, with a cloud at 0.1 bar, needed to agree with the \emph{HST} observations to 1$\sigma$ and 3$\sigma$. For planet g, the observational uncertainty is such that we are unable to generate any models that can be confidently excluded from agreement with the \emph{HST} data.}
\label{table:results}
\end{table}

\subsection{Haze} 

Adding a global layer of Rayleigh-scattering haze weakens spectral features short of 1.7 microns, as seen in our transmission spectra models in Figure \ref{fig:haze}. We increased the strength of the haze scattering cross section while maintaining a Rayleigh scattering slope of $\lambda$\textsuperscript{-4} as described in \S2.2. We determined that to 1$\sigma$ for TRAPPIST-1 d, e, and f, we can rule out models with large haze scattering cross sections. However, to 3$\sigma$, our model outputs with updated mass constraints cannot exclude purely haze-free solar composition atmospheres except for planet d.  

For TRAPPIST-1 d, a clear solar atmosphere is excluded to $>$20$\sigma$, while a clear hydrogen-rich atmosphere with 1\% water, carbon dioxide, nitrogen, and methane is excluded to $>$15$\sigma$. We are able to exclude these cases for TRAPPIST-1 d to such high certainty because of the planet's low gravity and (relatively) high temperature. At TRAPPIST-1 d's equilibrium temperature of 288 K, mass of 0.297 M$_\earth$, radius of 0.784 R$_\earth$, assuming a solar composition atmosphere with $\mu$ of 2.32, we obtain a scale height of 216 km. This is likely unphysical for such a small planet; for reference, Earth and Venus have scale heights of $\sim$8.5 km and 16 km, respectively. To 3$\sigma$, we exclude haze scattering cross sections of less than 6$\times$10$^{-20}$ cm$^2$ for TRAPPIST-1 d; to 1$\sigma$ we can increase the haze cross section up to 1$\times$10$^{-19}$ cm$^2$. This haze scattering cross section for TRAPPIST-1 d suppresses the molecular features of the spectrum to the point where only the small scattering slope remains to be observed. For reference, our computed cross sections for TRAPPIST-1 d to 1$\sigma$ and 3$\sigma$ are both on the order of 10$^{7}$ times that of the scattering of Earth's atmosphere.

For TRAPPIST-1 e, we can only exclude a clear solar atmosphere to 2.5$\sigma$, in contrast to the result of \citet{dewit2018atmospheric} which used previous mass estimates in their analysis to exclude such a case to high confidence ($>6\sigma$). A clear hydrogen-rich atmosphere with 1\% volatiles, however, is excluded with 5$\sigma$ certainty. We can rule out haze scattering cross sections of less than 9$\times$10$^{-25}$ cm$^2$ to 3$\sigma$ and 3$\times$10$^{-23}$ cm$^2$ to 1$\sigma$. These values are $\sim$ 450$\times$ and 14000$\times$ Earth's mean atmospheric scattering.

For TRAPPIST-1 f, a clear solar atmosphere, a 1\% volatiles atmosphere, or an atmosphere with a 3$\times$10$^{-25}$ cm$^2$ haze scattering cross section are unable to provide a solid 3$\sigma$ exclusion by the \emph{HST} data, with confidence values of 2$\sigma$, 2.8$\sigma$, and 2.5$\sigma$ respectively. For the hazy model, we maximize our confidence value at this haze scattering cross section of 3$\times$10$^{-25}$ cm$^2$ (125$\times$ Earth scattering). For our 1$\sigma$ cutoff, we are able to impose a maximum value of 6$\times$10$^{-23}$ cm$^2$ (28000$\times$ Earth scattering) for the haze scattering cross section.  

For TRAPPIST-1 g, the \emph{HST} observations are not sufficiently precise to exclude a clear 1\% volatile or solar composition atmosphere, as in \citet{dewit2018atmospheric}. As such, our hazy models do not provide any meaningful additional limits on the TRAPPIST-1 g atmosphere and will await future observations of higher precision. 

\subsection{Clouds}

Our cloudy atmosphere model results are displayed in Figure \ref{fig:clouds}. We are able to rule out a clear solar composition atmosphere for TRAPPIST-1 d. However, such clear atmospheres cannot be ruled out to high confidence for planets e, f, and g (the statistical significance of these cases are 2.5$\sigma$, 2$\sigma$, and $<1\sigma$, respectively). This result for planets e and f is in contrast to \citet{dewit2018atmospheric}, whose analysis depended on previous planet mass estimates.

\subsubsection{Increasing the Water Mixing Ratio with Tropospheric Clouds}
\begin{figure*}[h!]
\centering
\gridline{\fig{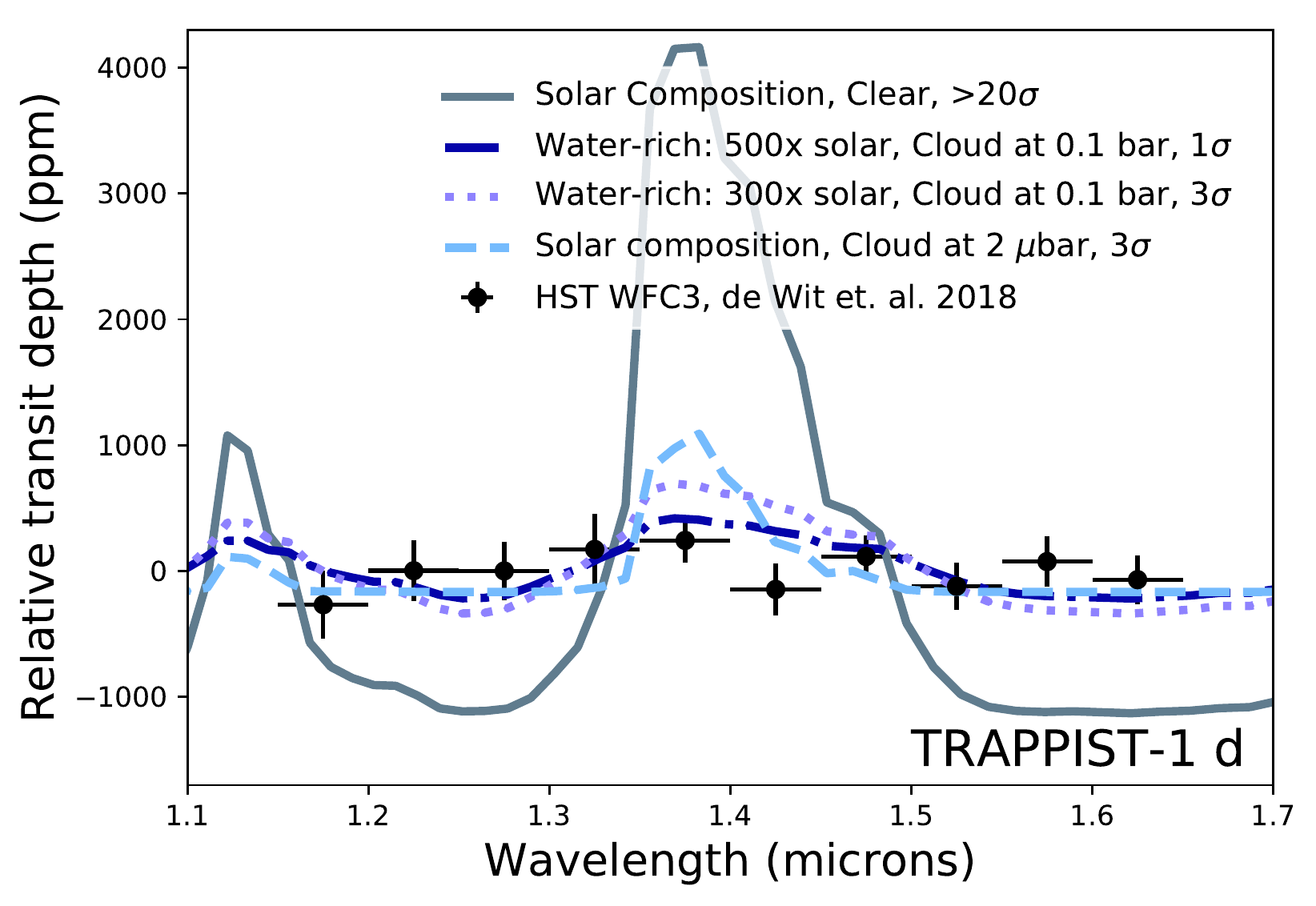}{0.5\linewidth}{}
          \fig{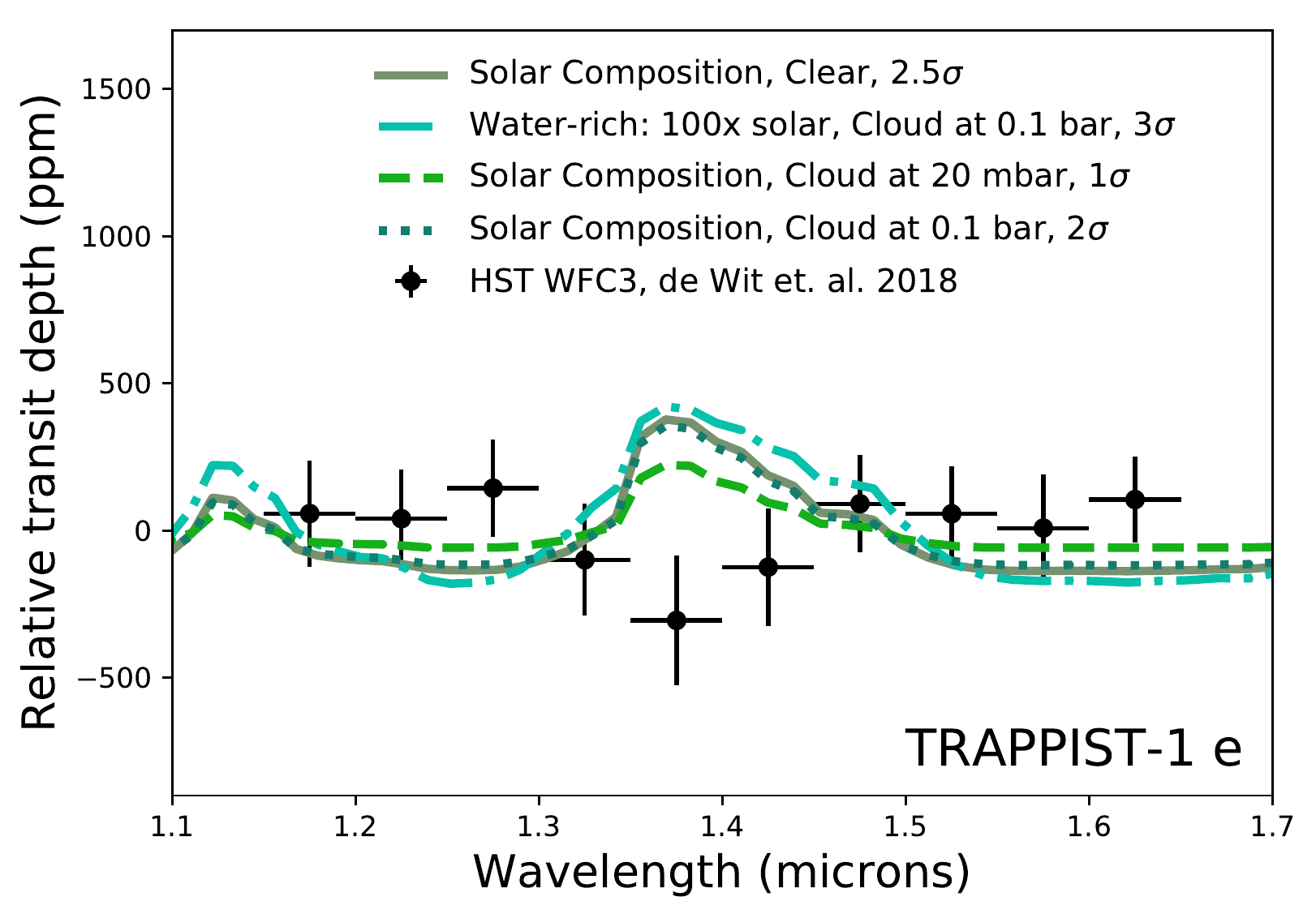}{0.5\linewidth}{}}
\gridline{\fig{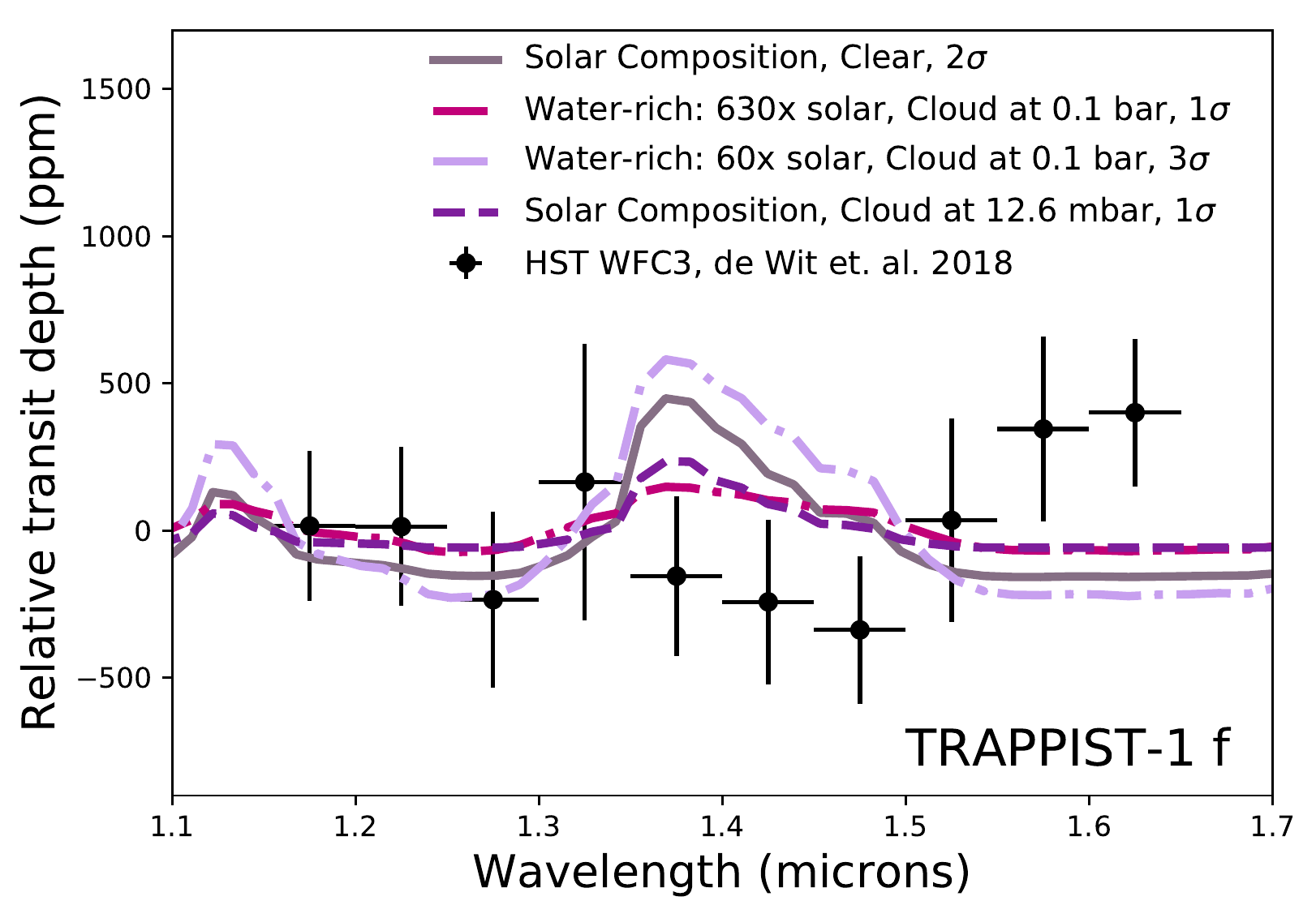}{0.5\linewidth}{}
          \fig{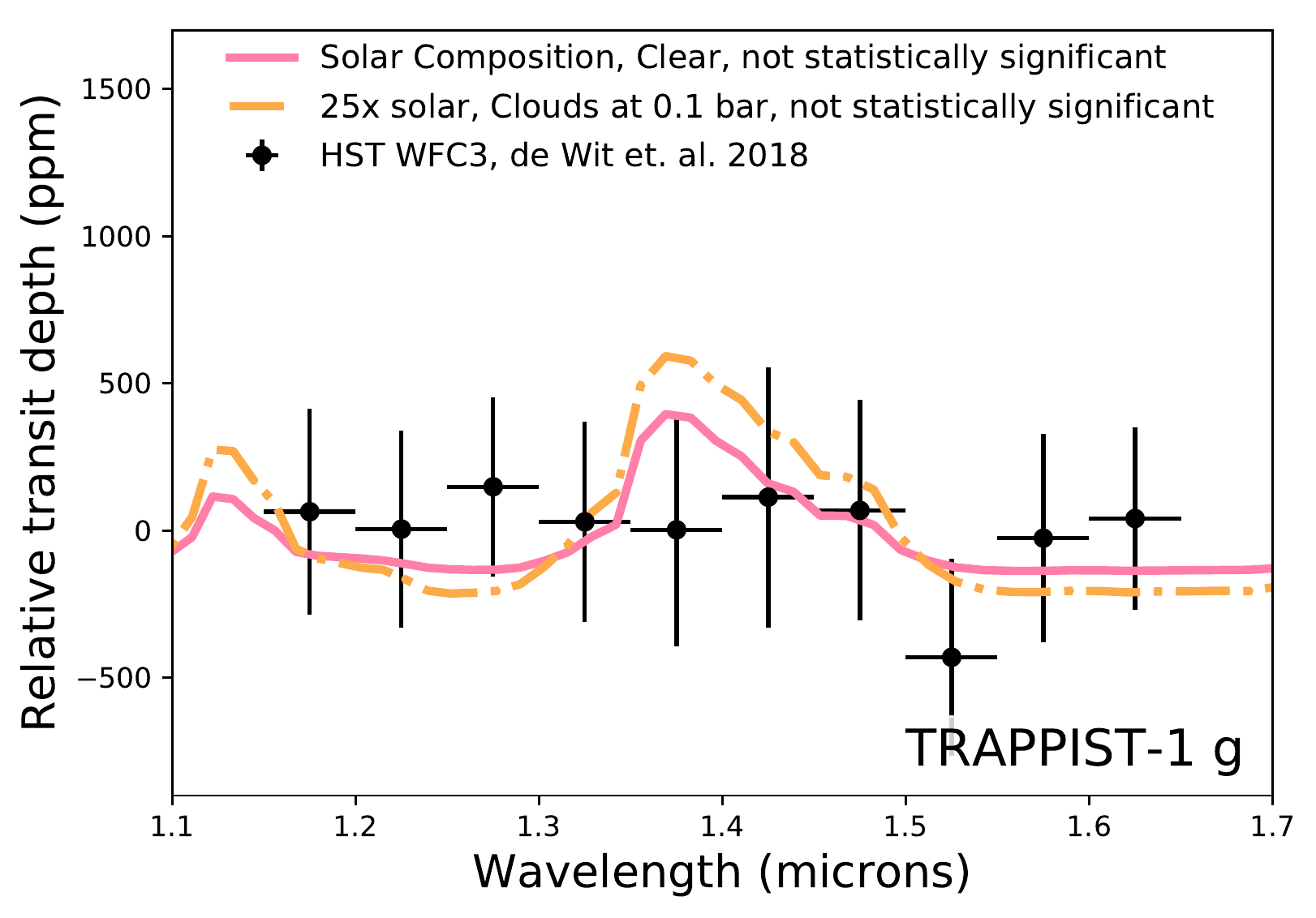}{0.5\linewidth}{}}
\caption{Black circles and error bars are from the \emph{HST}  observations \citep{dewit2018atmospheric} in all plots. Solid lines are model output atmospheres of solar composition with no clouds, dashed lines are for solar composition atmospheres with a cloud at high altitude, dash-dot lines are model outputs for metallicity-enhanced atmospheres with a cloud layer at the tropopause. All cases are labeled with statistical significance. We have chosen, in cases where multiple cases are statistically significant to the same confidence level, to show the higher metallicity value. Metal-rich atmospheres offer 1$\sigma$ agreement to the data for planets d, e, and f. High altitude clouds are required with planet d observations to both 1$\sigma$ and 3$\sigma$, but high clouds are only needed for planets e and f within the 1$\sigma$ uncertainty bound.} \label{fig:clouds}
\end{figure*}

With a grey cloud at 0.1 bar (the tropopause of Solar System bodies, following \citet{robinson2014tropopause}), the mean molecular weight must be supersolar in order to statistically match the \emph{HST} observations. This is shown in Figure \ref{fig:clouds}. For TRAPPIST-1 d and e, the model with a cloud deck at the tropopause requires an H$_2$O mixing ratio of 39\%, or $\sim$500$\times$ the solar ratio, to agree with the \emph{HST} observational error bars to 1$\sigma$. To 3$\sigma$, we calculate a water mixing ratio of 24\%, or a $\sim$300$\times$ solar atmosphere for TRAPPIST-1 d. For TRAPPIST-1 e, the tropospheric cloud layer necessitates a model with an 8\% water mixing ratio, or $\sim$100$\times$ solar, to 3$\sigma$ confidence. To produce a model with a cloudy tropopause, consistent at 1$\sigma$ with TRAPPIST-1 f observations, entails a 49\% water mixing ratio ($\sim$630$\times$ that of solar). At 3$\sigma$ confidence, however, only a 4.5\% water mixing ratio (60$\times$ solar) is needed. As in our hazy model results, adding clouds to the tropopause of a hydrogen-rich version of planet g offers no better statistical certainty than a clear hydrogen-rich atmosphere.

\subsubsection{Moving Clouds in Pressure Space}
If instead the cloud layer is allowed to move in pressure space while keeping the H$_2$O ratio steady at 0.08\% (a 1$\times$ solar ratio), we find that high altitude clouds ($<12 mbar$) are needed in the model for a 1$\sigma$ exclusion of the data except for planet g, where again no statistically significant model can be found with the current observational uncertainties. For TRAPPIST-1 d, e, and f, these clouds are at 0.8 $\mu$bar, 20 mbar, and 12.6 mbar, respectively. These pressures in Earth's atmosphere are comparable to that of the thermosphere (0.8 $\mu$bar) and stratosphere (20 and 12.6 mbar). Under our more conservative 3$\sigma$ cutoff, we rule out clouds below 2 $\mu$bar for TRAPPIST-1 d, again placing an opaque cloud deck in what would be the thermosphere on Earth. For TRAPPIST-1 e, we can only rule out clouds below the level of Earth's tropopause (0.1 bar) to 2$\sigma$ in a solar composition atmosphere while for f we cannot rule out a clear solar atmosphere beyond 2$\sigma$.
	
\subsubsection{The Intersection of Cloud Top Pressure and Atmospheric Metallicity}
\begin{figure*}[h!]
\centering
\includegraphics[angle=0,width=0.99\linewidth]{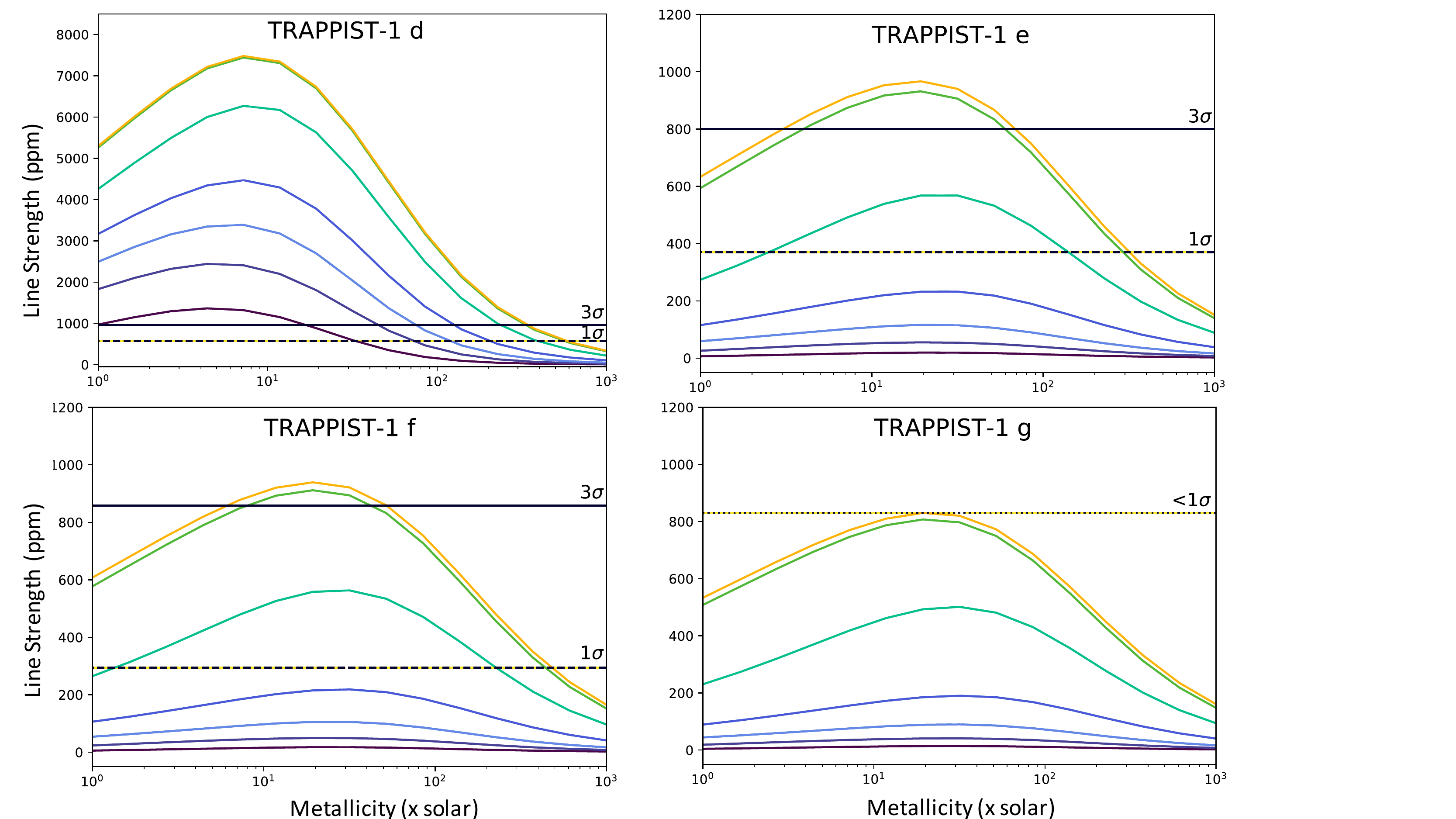}
\gridline{\fig{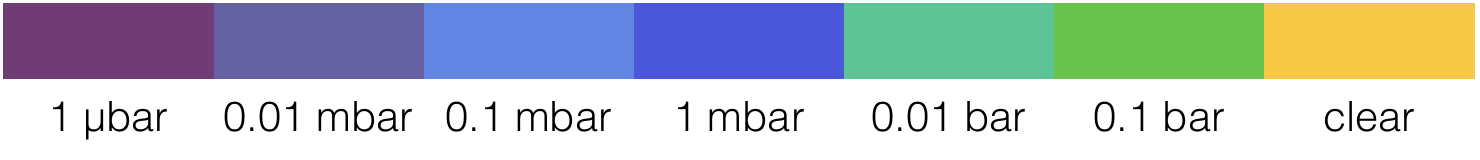}{0.6\linewidth}{}}
\caption{All plots show line strength as a function of metallicity for TRAPPIST-1 d, e, f and g (labeled). We define line strength as the difference between the maximum peak and minimum continuum of the transmission spectrum between 1.1 $\micron$ and 1.7 $\micron$. Each curve shows cloud-top pressures according to the color bar. 1$\sigma$ and 3$\sigma$ lines show the uncertainty bounds of the model to the \emph{HST} observations of \citet{dewit2018atmospheric}. For planet g, we plot the highest reduced-$\chi^2$ we were able to obtain as the line labeled $<$1$\sigma$. Note that planet d has a larger range of line strengths, due to its higher temperature and lower density coupling to give it a significantly larger scale height.} \label{fig:statistics}
\end{figure*}

There is a known degeneracy between cloud top pressure and metallicity, which both act to mute spectral features \citep{kempton2017exo,batalha2017challenges}. Here we explore both cloud top pressure level and metallicity, using the water mixing ratio as a metallicity proxy \citep[e.g.][]{batalha2017challenges}. We map the statistical significance from our previous analysis onto this parameter space. Figure \ref{fig:statistics} shows our results for these combined parameters. We define line strength as the difference between the maximum peak and minimum of the continuum of the transmission spectrum between 1.1 $\micron$ and 1.7 $\micron$. The line strength peak observable in the cloud contours results from the water feature at 1.4 $\mu$m. As the water content of the atmosphere increases, this water feature gets stronger and creates a spectral feature of larger amplitude. At the same time, increasing the water mixing ratio increases the mean molecular weight $\mu$. As scale height is an inverse function of mean molecular weight, adding more water decreases the scale height and thus weakens the line strength observed in transmission. A competition thus develops between scale height and the height of the individual water feature as water content increases, which manifests as the turnover observed in our contour plots.

 We can rule out distinct combinations of both metallicity and cloud top pressure for TRAPPIST-1 d, e, and f. For TRAPPIST-1 d, we can exclude clouds at 1$\mu$bar up to 20$\times$ solar metallicity, and a clear atmosphere up to around 400$\times$ solar. For TRAPPIST-1 e, we can exclude ``tropospheric'' clouds at 0.1 bar between 4$\times$ solar and 100$\times$ solar, while for TRAPPIST-1 f we exclude clouds at 0.1 bar between 8$\times$ solar and 60$\times$ solar. Again, as the precision of observational data for TRAPPIST-1 g is relatively poor, we are unable to make any significant diagnostics across either metallicity or cloud pressure space. Additionally, we show the level of precision needed to distinguish between various high metallicity atmospheres with clouds for all planets, regardless of the current \emph{HST} observations. For example, we show for TRAPPIST-1 d that the difference between a clear atmosphere and a cloud deck at the tropopause will not be observable even with 1 ppm precision. For TRAPPIST-1 e and f, we will be able to distinguish between a cloudy tropopause and a completely clear atmosphere at approximately 20 ppm precision.

\section{Discussion} \label{sec:discussion}

\subsection{Stellar Contamination in the Transmission Spectra}
Our results are predicated on the observations of \citet{dewit2018atmospheric}. \citet{rackham2018transit} recently called into question the fidelity of these \emph{HST} measurements because of the effects of unaccounted-for stellar contamination. Spots and faculae in the photosphere of M-dwarf host stars, such as TRAPPIST-1~A, may contribute to stellar contamination in the planetary spectra, which \citet{rackham2018transit} refer to as the ``transit light source effect''. Specifically, this effect impacts the near-IR \emph{HST}/WFC3 bandpass to which we compare our models. To address this question of stellar contamination, \citet{zhang2018nearinfrared} reanalyze the \emph{HST}/WFC3 data previously published by \citet{dewit2018atmospheric}. They use a different data reduction strategy in an attempt to minimize systematics, and ultimately find transit depths consistent with those of \citet{dewit2018atmospheric} within error bars. However, they  also model the ``transit light source effect'' and determine that these data are fully consistent with stellar contamination in the transmission spectrum.

With this possibility of stellar contamination in the TRAPPIST-1 spectra \citep{zhang2018nearinfrared},  \citet{rackham2018transit} suggest that any molecular features in the region of interest (as here, from 1.1 to 1.7 \micron) would be impacted up to 77 ppm. However, for the \emph{HST}/WFC3 data presented in \citet{dewit2018atmospheric}, the uncertainty due to signal and potentially unaccounted-for instrument systematics are actually larger than this across the entire wavelength band, with a minimum error twice that ($\sim$145 ppm). Only for higher precision transit observations, such as with the \emph{James Webb Space Telescope (JWST)} with a noise floor of 30 ppm \citep{batalha2018strategies}, is stellar contamination likely to impact planetary spectra and molecular feature identification. Still, in light of these complexities, our analysis, as described in \S2, does not attempt to \emph{fit} our model to the \emph{HST} data. Instead, in order to calculate conservative limits of the metallicity, cloud top pressure, and haze scattering cross sections of the TRAPPIST-1 planets d, e, f, and g, we only report agreement of our models within the \emph{HST} uncertainty bounds. Our analysis presented here is therefore minimally affected by spots or faculae in the stellar spectrum. Our results do not depend on a true fitting of the transit spectra and further efforts to characterize any stellar contamination present would likely only reduce the uncertainties inherent to the \emph{HST} data and improve our results.

\subsection{Effect of Temperature-Pressure Profiles and Planetary Mass on Scale Heights}
We use water as a proxy for increased metallicity in our model atmospheres (see \S2). We find that for TRAPPIST-1 d, e, and f, 500$\times$ to 600$\times$ solar metallicities are within the bounds of the \emph{HST} uncertainties to 1$\sigma$ if a tropospheric cloud layer is included in the model. These results imply large ($>$ 6) mean molecular weights, $\mu$, which work to reduce the scale height of the atmosphere, given by $H = \frac{kT}{\mu g}$. However, the effect of temperature \emph{T} on scale height is dependent on our assumed T-P profiles, where we use the parameterized 1-D, 5 parameter profile of \citet{guillot2010TPprofile}. Our T-P profiles are in agreement with those explored in \citet{morley2017observing} and \citet{batalha2018strategies}; however the true temperature structures of the TRAPPIST-1 planets are highly unconstrained. If the true temperatures were warmer, our upper limits for metallicity would have to be larger to compensate for this increase in temperature. Alternatively, our models would have to include higher altitude clouds or more strongly scattering haze particles. Quantitatively, any 50 K difference in our models changes our upper limit of metallicity by $\sim$2-2.5 dex (e.g., from 500 to 700$\times$solar) for planet d and 1 dex for planets e and f. Our cloud top pressure changes by 1 mbar for planet d, 10 mbar for planet e, and 6 mbar for planet f with a 50 K change in temperature. Finally, our haze scattering cross sections change by no more than 9$\times$10$^{-27}$ cm$^2$ for all planets with any 50 K temperature adjustment. This 50 K change is within the range of albedo explored by \citet{morley2017observing}. Because we only consider transmission spectra, as opposed to highly temperature-dependent emission spectra, our results are relatively unchanged by our choice of T-P profile parameterization.   With our current T-P parameterization, our results provide a conservative limit to the metallicity and/or aerosol content of the TRAPPIST-1 planet atmospheres.

Like temperature, the mass of the planet is linearly related to the scale height. The masses of the planets in the TRAPPIST-1 system have undergone several refinements with additional TTV measurements, as reported in \citet{grimm2018nature}. These mass updates changed the mass estimates by 10\% to 25\%, and correspondingly affect the computed scale heights for our model atmospheres by the same amount. This allows us to show that the solar composition atmospheres which \citet{dewit2018atmospheric} exclude for TRAPPIST-1 e and f cannot be discounted. Any further mass refinements will thus affect the atmospheric scale heights and could substantially change the atmospheric metallicity and aerosol contents we find in this work. For example, if the planetary mass of TRAPPIST-1 e were to decrease by 15\%, we could exclude a hydrogen-rich atmosphere to 3$\sigma$. If TRAPPIST-1 d were to increase in mass by 25\%, our exclusion of a hydrogen atmosphere would fall from $>$20$\sigma$ to only $>$7$\sigma$. These examples demonstrate the extreme importance of having high-precision, high-fidelity mass measurements for these small planets, as determining the nature of their atmospheres is highly dependent on this information.

\subsection{Aerosol Mass Loading}
For our cloudy models, our findings suggest a solid layer of high altitude clouds for TRAPPIST-1 d, e, and f as an upper limit within 1$\sigma$. The lower boundary of the cloud layers are, at minimum, at 0.8 $\mu$bar, 20 mbar, and 12.6 mbar respectively for each planet, consistent with pressures in the thermosphere (0.8 $\mu$bar) and stratosphere (20 and 12.6 mbar) on Earth. However, our treatment of the cloud opacity does not attempt to self-consistently model the cloud formation. In a real atmosphere, there is likely not enough material for solid, grey clouds to form at the altitudes of our lower limits -- stratospheric clouds and higher are optically thin on Earth due to the low number density of molecules available at these millibar pressures \citep{seinfeldtextbook}. A solid grey cloud at the 0.01 bar level is the most pessimistic case considered in \citet{batalha2018strategies}, for example. However, \citet{kopparapu2017habitable} suggest that for slowly or synchronously rotating planets, thick convective clouds may be more easily able to form and persist at higher altitudes. Despite this, for TRAPPIST-1 d, even at 3$\sigma$, an opaque cloud no lower than 2 $\mu$bar is required by our model. This pressure can be safely ruled out even in the most generous of cloud formation models, implying that clouds are not the source of the observed muted transmission features for planet d. At 3$\sigma$ only TRAPPIST-1 e allows a cloud at 0.01 bar, but e has a density of 1.024 $\rho_\earth$ \citep{grimm2018nature}, which is consistent with a volatile-rich rather than extended hydrogen-rich atmosphere \cite[]{lopez2014composition,rogers2015rocky}.

If we turn to haze scattering as an aerosol source, we find haze scattering cross sections 10$^2$ to 10$^7$ times that of Earth's atmospheric scattering are needed to be within the uncertainty of the \emph{HST} observations for planets TRAPPIST-1 d, e, and f. However, laboratory results show that hydrogen-rich atmospheres are not very efficient at making haze \citep{he2018uvhaze,horst2018haze}, and all of our model atmospheres considered here are heavily hydrogen-rich with minor volatile contents. For Titan, the haze scattering cross section is $\sim$10$^{-7}$ cm$^2$, as described in \S2.1. Laboratory results for Titan's haze production rate are the same order of magnitude for the most efficient exoplanet haze production rates \citep{horst2018haze}, suggesting that any haze scattering cross sections substantially greater than Titan's are unphysical. However, the most haze-productive laboratory atmospheres were run at temperatures of 400 K (the equilibrium temperature of TRAPPIST-1 b rather than those of the outer planets d, e, f, and g considered here). Furthermore, these highly productive laboratory experiments contained more metal-rich atmospheric compositions (1000$\times$ solar) than our models ($\sim$100$\times$ solar). For the laboratory atmospheres comparable to TRAPPIST-1 d conditions considered here ($~$300 K, $\sim$100$\times$ solar), the haze production rate is three orders of magnitude lower. Again, there is no direct way to translate laboratory-measured haze production rates to haze mass loading in a planetary atmosphere. Still, it is likely that the less productive laboratory cases also represent less hazy worlds. Our results only conclusively rule out haze scattering cross sections well under that of Titan-like conditions, but the disconnect between the laboratory conditions and our models means that more precise constraints remain elusive.

\citet{lincowski2018trappist} used a coupled photochemical, climate, and radiative transfer model to consider the effects of Earth- and Venus-like aerosols such as water and sulfuric acid clouds as well as the photochemical products of these and other molecules within the TRAPPIST-1 planets, for oxygen-rich and carbon dioxide-rich atmospheres. They found aerosol scattering cross sections up to 10$^3$ times that of Earth's ozone, which is consistent with our result for the upper limit of the haze scattering cross section suggested by the laboratory exoplanet haze samples. Within the current \emph{HST} observational uncertainty, we cannot make statistically significant determinations regarding the likelihood of such hazy metal-rich atmospheres based on observations. Our findings thus represent an upper limit to haze in hydrogen-rich models of the TRAPPIST-1 outer planet atmospheres. The laboratory measurements of \citet{horst2018haze} and \citet[]{he2018laboratory,he2018uvhaze}, as well as the modeling work of \citet{lincowski2018trappist}, help to inform whether our results are physically realistic. More observations to better precision of the TRAPPIST-1 planets (as are currently planned with \emph{JWST} GTO Cycle 1) will be required to make further predictions as to the haze content of these worlds. Figure \ref{fig:statistics} shows the precision required to rule out aerosols in heavy mean molecular weight atmospheres. For TRAPPIST-1 e and f, the worst-case scenario of 1000$\times$ solar with low clouds requires a precision of approximately 20 ppm, which is beyond the expected 30 ppm noise floor for \emph{JWST} \citep{batalha2018strategies}. However, for more optimistic scenarios, 50 to 100 ppm would be enough to differentiate between various cloud cases for metallicities on the order of 100$\times$ solar.

\subsection{Complexity of Combined Parameters}
In Figure \ref{fig:statistics}, we show the intersection of cloud top pressure levels with increasing metallicity and how this changes the observed strength of the transmission spectra. These results demonstrate the important likelihood that a combination of factors is at play in the small ($<$ 500 ppm) features of the TRAPPIST-1 spectra. If indeed these atmospheres are not mainly primordial hydrogen, but secondary and composed of higher metallicity species, we may begin to speculate as to the types of clouds present in these atmospheres and on the ability of these cloud species to form thick, grey absorbing clouds. \citet{schaefer2012vaporization} suggests that H$_2$O and CO$_2$ are the likeliest components of secondary atmospheres of Earth-like planets, and therefore these cloud species merit further investigation \citep{marley2013clouds}.

\subsection{Aerosol Particle Properties}
While we include in our models the effects of Rayleigh scattering haze, this does not capture the complexity of the full distribution of particles that may exist in the TRAPPIST-1 planet atmospheres. The laboratory results upon which we base our models show that very small particles are produced and are readily treatable by the Rayleigh approximation at \emph{HST}/WFC3 wavelengths. However, it is possible that in a real planetary atmosphere, haze particle aggregates would form and grow large enough that their effect on radiative transfer would lie in the Mie regime (where the particle diameter d$_p$ $\sim$ $\lambda$) and require full treatment with Mie theory to capture \citep[e.g.,][]{wakeford2015particles,kitzmann2018optical}. 

Additionally, the optical properties of particles must still be accounted for even if particles are small. In fact, the laboratory results of \citet{he2018laboratory} show that the exoplanet haze particles have varying colors in the visible, which may suggest that their scattering and absorption properties in the near-IR wavelengths considered here may not be a simple matter of inducing a Rayleigh scattering slope. This question awaits further laboratory measurements to characterize the optical properties of these haze analogues, which can then be more rigorously implemented into our models.

\subsection{Future Observations}
While the current \emph{HST} observations have considerable limitations that prevent robust, specific predictions of the atmospheric properties of such small planets as found in the TRAPPIST-1 system, the \emph{James Webb Space Telescope (JWST)} will have both the resolution and wavelength coverage to greatly enhance our ability to measure their atmospheres \citep{batalha2018strategies}. The effects of differing atmospheric compositions as well as the effects of any aerosols should be observable for several of the TRAPPIST-1 planets in only a few orbits \citep{morley2017observing}. Furthermore, upcoming \emph{HST} observations of TRAPPIST-1 g (GO Proposal 15304, PI J. de Wit) may offer additional precision on its atmosphere. Planet h's atmosphere has not yet been observed in transit, though these observations are also upcoming (GO Proposal 15304, PI J. de Wit), and would naturally provide additional information as to the nature of this system. Finally, our results show that, in light of updated mass measurements, the previous \emph{HST} observations do not rule out hydrogen-rich atmospheres for either planet e or f, as found by \citet{dewit2018atmospheric}. This motivates new observations with higher precision, such as \emph{JWST} can achieve, to provide better constraints on these atmospheres.

\section{Conclusion} \label{sec:conclusions}

We have performed a modeling analysis supported by recent laboratory measurements to explore the nature of the outer TRAPPIST-1 planetary atmospheres. We find that, using laboratory-based and Solar System constraints for haze formation, there are upper limits on haze scattering cross sections in a hydrogen atmosphere to high statistical certainty with the \emph{HST} data. These haze scattering cross sections range from a minimum of 9$\times$10$^{-25}$ cm$^2$ for planet e to a maximum of 1$\times$10$^{-19}$ cm$^2$ for planet d. We found a minimum and maximum metallicity for a case with a 0.1 bar grey opacity source. Using water as a proxy for metallicity, we find that planets d, e, and f allow, at maximum, hydrogen-to-water mixing ratios of 500 to 630$\times$ solar, respectively. For our cloudier cases, we find that a high altitude cloud deck (12 mbar or lower in pressure) is needed to generate a model within the current precision of the \emph{HST} data for planets d and e with solar composition. It is likely unphysical that such clouds could form and persist in the TRAPPIST-1 d and e atmospheres. The possibility of enhanced atmospheric metallicity has also been posed by previous studies about the water content of the TRAPPIST-1 system. Some of our results differ considerably from those of the original \emph{HST} analysis; this difference 
ensues from additional mass constraints of the TRAPPIST-1 planets. High-precision mass measurements are of utmost importance to constrain the atmospheres of small terrestrial planets, and any further improvements on mass will allow better estimates of both atmospheric composition and aerosol content.

Our results further support secondary, post-primordial atmospheres for the TRAPPIST-1 planets d, e, and f, which could include substantial amounts of aerosols. Here we seek only to provide limits on the possible metallicity, cloud top pressure, and haze scattering cross sections of these atmospheres in light of the recent \emph{HST} campaign. While our results suggest that the outer worlds d, e, and f of the TRAPPIST-1 system could have volatile-rich secondary atmospheres, determining the aerosol content of such volatile-rich atmospheres requires greater precision than the current set of \emph{HST} data can provide. We show that at least 20 ppm precision will be needed to discern between cloudy versus clear cases in high metallicity atmospheres. Further investigations into the habitability of these worlds must include full consideration of atmospheric composition and aerosol content. In light of these possibilities, the TRAPPIST-1 planets should be of high priority for further examination with both current and future observatories. 

\acknowledgments
The authors thank Julien de Wit for his support on this project and Michael R. Line for use of the \texttt{CHIMERA} model. We also thank the STScI STARGATE team for all of their helpful discussion. We thank the H\"{o}rst PHAZER lab group at JHU EPS for their useful discussion and eagle-eyed editing expertise. S.E.M. thanks J.M., Z.P.-M., and S.P.-M. for their support during the writing of this manuscript. This study was supported in part by a Hubble Grant associated with program GO-14873. S.E.M. was supported in part by a Johns Hopkins University Catalyst Award.

\bibliography{apjmnemonic,clouds}
\bibliographystyle{aasjournal}



\end{document}